\documentclass[12pt, draftclsnofoot, journal, letterpaper, onecolumn]{IEEEtran}

\usepackage[dvips]{graphicx}
\usepackage{times}
\usepackage{cite}
\usepackage{amsmath}
\usepackage{array}
\usepackage{amssymb}

\usepackage{stfloats}
\usepackage{rotating,threeparttable,booktabs}
\usepackage{bm}
\usepackage{dcolumn,booktabs}
\begin{document}

\title{Blind Spectrum Sensing by Information Theoretic Criteria for Cognitive Radios}

\author{{Rui~Wang and Meixia~Tao}
\thanks{The authors are with the Department of Electronic Engineering at Shanghai Jiao Tong University, Shanghai, 200240,
P. R. China. Emails:\{liouxingrui, mxtao\}@sjtu.edu.cn}}

\maketitle


\begin{abstract}
Spectrum sensing is a fundamental and critical issue for
opportunistic spectrum access in cognitive radio networks. Among the
many spectrum sensing methods, the information theoretic criteria
(ITC) based method is a promising blind method which can reliably
detect the primary users while requiring little prior information.
In this paper, we provide an intensive treatment on the ITC sensing
method. To this end, we first introduce a new over-determined
channel model constructed by applying multiple antennas or over
sampling at the secondary user in order to make the ITC applicable.
Then, a simplified ITC sensing algorithm is introduced, which needs
to compute and compare only two decision values. Compared with the
original ITC sensing algorithm, the simplified algorithm
significantly reduces the computational complexity without losing
any performance. Applying the recent advances in random matrix
theory, we then derive closed-form expressions to tightly
approximate both the probability of false alarm and probability of
detection. Based on the insight derived from the analytical study,
we further present a generalized ITC sensing algorithm which can
provide flexible tradeoff between the probability of detection and
probability of false alarm. Finally, comprehensive simulations are
carried out to evaluate the performance of the proposed ITC sensing
algorithms. Results show that they considerably outperform other
blind spectrum sensing methods in certain cases.

\end{abstract}


\section{Introduction}
\setlength\arraycolsep{2pt}
Due to the increasing popularity of wireless devices in recent years,
the radio spectrum has been
an extremely scarce resource. By contrast,
90 percent of the existing licensed spectrum remains idle
and the usage varies geographically and temporally
as reported by the Federal Communication Commission (FCC) \cite{fcc2002}.
This indicates that the fixed frequency regulation policy conflicts drastically with the high demand
for frequency resource. Cognitive radio (CR) is one of the most promising technologies
to deal with such irrational frequency regulation policy \cite{Haykin05, Zhao2007a} and
has received lots of attention. In cognitive radio networks, secondary (unlicensed) users
first reliably sense the primary (licensed) channel and then opportunistically
access it without causing harmful interference to primary users \cite{fcc2008}. By doing this,
the spectrum utilization of existing wireless communication networks can be tremendously improved.
FCC has issued a Notice of proposed Rule Making to allow the unlicensed CR devices to operate in the
unused channel \cite{fcc2003}. The IEEE has also formed the 802.22 working group to develop the standard
for wireless regional area networks (WRAN) which will operate on unused VHR/UHF TV bands based on cognitive
radio technology. Both of these activities will significantly change the current wireless communication situation.

As mentioned above, the secondary users need to opportunistically
access the unused licensed channel while causing negligible
interference to the primary users. As a result, the detection of
presence of primary users is a fundamental and critical task in the
cognitive radio networks. Although the detection of presence of
signals is known as a classical problem in signal processing,
however, sensing the presence of primary users in a complicated
communication environment, especially a CR-based network, is still a
challenging problem from the practice perspective. This is mainly
due to the following two limiting factors: Firstly, it is very
difficult, if not possible, for the secondary user to obtain the
necessary prior information about the signal characteristics of the
primary user for most of the traditional detection techniques to
apply. Secondly, the CR devices should be capable of sensing the
very weak signals transmitted by primary users. For instance, the
standard released by FCC has required that spectrum sensing
algorithms need to reliably detect the transmitted TV signals at a
very low signal-to-noise ratio (SNR) of at least $-18$dB
\cite{fcc2008}.

Thus far, there are mainly four types of spectrum sensing methods: energy detection \cite{Urkowitz1967, Digham2007},
matched filtering (coherent detection) \cite{Hwang09}, feature detection \cite{Danda94}
and eigenvalue-based detection \cite{Zeng09,zeng08sp,RuiZhang10com}.
Among them, energy detection is optimal if the secondary user only knows the local noise power\cite{kay98}.
The matched-filtering based coherent detection is optimal for maximizing
the detection probability but it
requires the explicit knowledge
of the transmitted signal pattern (e.g., pilot, training sequence etc.) of the primary user. The feature detection,
often referred to as
the cyclostationary detection, exploits the periodicity in the modulation scheme which, however, is difficult to determine in certain scenarios.
By constructing the decision variables based on
eigenvalues of the sampled covariance matrix to detect the presence of the primary user,
the eigenvalue-based sensing methods presented in \cite{Zeng09,zeng08sp,RuiZhang10com} do not need to estimate the power of the noise and hence are more practical in most CR networks.
Recently, several new spectrum sensing schemes by incorporating system-level design parameters have been introduced,
such as throughput maximization \cite{Liang08, Quan09, Shen09} and cooperative sensing using multiple nodes
\cite{Ganesan08, Ganesan07I, Ganesan07II, Jun08}.
Nevertheless, the aforementioned four types of sensing techniques are still treated as
a basic component in these new schemes.

In this paper, we study a blind spectrum sensing method based on information theoretic criteria (ITC),
an approach originally for model selection introduced by Akaike \cite{Akike73,
Akaike74} and by Schwartz \cite{Schwartz78} and Rissanen \cite{Riss78}.
Applying information theoretic criteria for spectrum sensing was firstly introduced in \cite{Zayen09, Zayen08, Majed07, Zayen10}. This work provides a more intensive study on the ITC sensing algorithm and its performance. The main contributions of this paper are as follows:
\begin{itemize}
\item First of all, to make the information theoretic criteria applicable, a new over-determined channel model is constructed by introducing multiple antennas or over sampling at the secondary user.
\item Then, a simplified information theory criteria (SITC) sensing algorithm which only involves the computation of two decision values is presented. Compared to the original information theory criteria (OITC) sensing algorithm in \cite{Zayen09}, SITC is much less complex and yet almost has no performance loss. Simulation results  also demonstrate that the proposed SITC based spectrum sensing outperforms the eigenvalue based sensing algorithm in \cite{Zeng09} and almost obtains the similar performance with \cite{zeng08sp}. The proposed sensing algorithm also enables a more tractable analytical study on the detection performance.
\item Applying the recent advances in random matrix theory, we then derive closed-form expressions for both the probability of false alarm and probability of detection. which can approximate the actual results in simulation very well.
\item Finally, based on the insight derived from the analytical study, we further present a generalized information theory criteria (GITC) sensing algorithm. By involving an adjustable threshold, the proposed GITC can provide flexible tradeoff between the probability of detection and probability of false alarm in order to supply different system requirements.
\end{itemize}

The rest of paper is organized as follows. In Section II, the
preliminary on the information theoretic criteria is provided. The
proposed over-determined system model is presented in Section III.
Section IV gives the proposed SITC sensing algorithm and the theoretical analysis
of its detection performance, followed by the GITC sensing algorithm in Section V. Extensive
simulation results are illustrated in Section VI. Finally, Section
VII offers some concluding remarks.

\emph{Notations}: $\cal E[\cdot]$ denotes expectation over the
random variables within the brackets. ${\rm Tr}({\bf A})$ stands for the trace of matrix ${\bf A}$.
Superscripts $(\cdot)^T$ and $(\cdot)^\dagger$ denote transpose and conjugate transpose.

\section{Preliminary on the Information Theoretic criteria}
Information theoretic criteria are an approach originally
for model selection introduced by Akaike \cite{Akike73,
Akaike74} and by Schwartz \cite{Schwartz78} and Rissanen \cite{Riss78}.
There are two well-known criteria that have been widely used: Akaike
information criterion (AIC) and minimum description length (MDL)
criterion. One of the most important applications of information theoretic criteria is to estimate the number of
source signals in array signal processing \cite{Wax85}.
Consider a system model described as
\begin{equation} \label{eqn:AICMDLmodel}
    {\bm x}={\bf A}{{\bm s}}+{\bm{\mu}},
\end{equation}
where $\bm x$ is the $p\times 1$ complex observation vector, $\bf A$ is a $p\times q$ ($p>q$)
complex system matrix, $\bm s$ denotes the $q\times1$ complex source modulated signals and $\bm{\mu}$ is
the additive complex white Gaussian noise vector.
It is noted that the definite parameters $q$, $\bf A$ and $\sigma^2$ are all unknown.
The resulting cost functions of AIC and MDL have the following form \cite{Wax85}:
\begin{equation} \label{eqn:AICsolve}
    {{\rm AIC} (k)}={-2\log \left (\frac{\prod^p_{i=k+1}{l^{1/(p-k)}_i}}{{\frac{1}{p-k}}\sum^p_{i=k+1}{l_i}}\right )^{N(p-k)}}+2k(2p-k)+2,
\end{equation}
\begin{equation} \label{eqn:MDLsolve}
    {{\rm MDL} (k)}={-\log \left (\frac{\prod^p_{i=k+1}{l^{1/(p-k)}_i}}{{\frac{1}{p-k}}\sum^p_{i=k+1}{l_i}}\right )^{N(p-k)}}
    +\left({\frac{1}{2}}k(2p-k)+\frac{1}{2}\right)\log N,
\end{equation}
where $N$ signifies the observation times and $l_i$ denotes the ${i}$-th decreasing ordered eigenvalue of the sampled covariance
matrix.
The estimated number of source signals
is determined by choosing the minimum \eqref{eqn:AICsolve} or \eqref{eqn:MDLsolve}. That is,
\begin{equation} \label{eqn:AICestimate}
    {\hat k_{\rm AIC}}=\mathrm{arg} {\min_{j=0,1,\ldots,p-1}} {\rm AIC}(j),
\end{equation}
\begin{equation} \label{eqn:MDLestimate}
    {\hat k_{\rm MDL}}=\mathrm{arg} \min_{j=0,1,\ldots,p-1} {\rm MDL}(j).
\end{equation}

\section{System Model}
We consider a multipath fading channel model and assuming that there is only one primary user in the cogitative radio network.
Let $x(t)$ be a continuous-time baseband received signal at the secondary user's receiver.
Spectrum sensing can be formulated as a binary hypothesis test between the following two hypotheses
\begin{equation} \label{eqn:H0con}
    {{\cal H}_0:} \quad   x(t)=\mu(t),
\end{equation}
\begin{equation} \label{eqn:H1con}
    {{\cal H}_1:} \quad  x(t)=\int^{T}_0 {h(\ell)s(t-\ell)}{d\ell}+\mu(t),
\end{equation}
where $s(t)$ denotes the signal transmitted by the primary user,
$h(t)$ is the continuous channel response between the primary
transmitter and the secondary receiver, $\mu(t)$ denotes the
additive white noise, the parameter $T$ signifies the duration of
the channel.
The channel response is also assumed to remain invariant during each observation.
To obtain the discrete representation, we assume that
the received signal is sampled at rate $f_s$ which is equal to the
reciprocal of the baseband symbol duration $T_0$. For notation
simplicity, we define $x(n)=x(nT_0)$, $s(n)=s(nT_0)$ and
$\mu(n)=\mu(nT_0)$. Hence, the corresponding received signal samples
under the two hypotheses are described as:
\begin{equation} \label{eqn:H0dis}
    {{\cal H}_0:} \quad   x(n)=\mu(n),
\end{equation}
\begin{equation} \label{eqn:H1dis}
    {{\cal H}_1:} \quad  x(n)=\sum^{L-1}_{i=0}{h(i)s(n-i)}+\mu(n),
\end{equation}\
where $h(i)$ ($0 \leqslant i \leqslant L-1$) denotes the discrete channel response of $h(t)$
and $L$ denotes the order of the discrete channel ($L$ taps).
Let each observation consist of $M$ received signal samples.
Then \eqref{eqn:H0dis} and \eqref{eqn:H1dis} can be rewritten in
matrix form as:
\begin{equation} \label{eqn:H0mat}
    {{\cal H}_0:} \quad   {\bm x}_i=\bm \mu_i,
\end{equation}
\begin{equation} \label{eqn:H1mat}
    {{\cal H}_1:} \quad  {\bm x}_i={\bf H}{\bm s}_i+\bm \mu_i,
\end{equation}
where $\bf{H}$ is an $M\times(L+M-1)$ circular channel matrix defined as
\begin{equation} \label{eqn:ChannH}
              {\bf H}=\left[ \begin{array}{ccccccccc}
                      h(L-1) &  h(L-2) &  \ldots  &   h(0)\\
                       &        h(L-1) &   h(L-2) &   \ldots &  h(0)\\
                       &         &       \ddots &          &  \ddots \\
                       &         &          &          &         &     h(L-1) & h(L-2) & \ldots &  h(0)
                      \end{array} \right],\nonumber
\end{equation}
$\bm{x}_i$, $\bm{s}_i$, and $\bm{\mu}_i$ are the $M\times1$ observation vector, $(L+M-1) \times1$ source signal vector and $M\times1$ noise vector,
respectively, and are defined as
\begin{equation} \label{eqn:Xn}
   {{\bm x}_i}={[x(iM-M+1),x(iM-M+2),\ldots,x(iM)]^T},
\end{equation}
\begin{equation} \label{eqn:Sn}
    {{\bm s}_i}={[s(iM-M-L+2),s(iM-M-L+3),\ldots,s(iM)]^T},
\end{equation}
\begin{equation} \label{eqn:Nn}
    {{\bm \mu}_i}={[\mu(iM-M+1),\mu(iM-M+2),\ldots,\mu(iM)]^T}.
\end{equation}
Now, comparing \eqref{eqn:H1mat} with the array signal processing model \eqref{eqn:AICMDLmodel}, we
find that a major difference is that the $\bf H$ in our considered
system model is an under-determined matrix, i.e., the order of
column is larger than the order of row. Therefore, the information
theoretic criteria are not directly applicable here \cite{Wax85}.

To construct an over-determined channel matrix $\bf{H}$ as in \eqref{eqn:AICMDLmodel}, one needs to enlarge the observation space. Obviously, simply increasing the
observation window $M$ does not work. Here we propose to expand the observation space using one of the following two methods.
One is to increase the spatial dimensionality by employing multiple receive antennas at the secondary user and the other is to increase
the time dimensionality by over-sampling the received signals. It turns out that the two methods are similar to each other.
Hence we shall focus on the multiple-antenna approach hereafter.
The difference for over-sampling method will be discussed in the end of this section.
In specific, suppose that the detector at the secondary user is equipped with $K$ antennas.
Redefine \eqref{eqn:Xn} and \eqref{eqn:Nn} as
\begin{equation} \label{eqn:newXn}
    {{\bm x}_i}=[x^i_1(1),x^i_2(1),\ldots,x^i_K(1),x^i_1(2),\ldots,x^i_K(2),\ldots,x^i_1(M),\ldots,x^i_K(M)]^T,
\end{equation}
\begin{equation} \label{eqn:newNn}
    {{\bm \mu}_i}=[\mu^i_1(1),\mu^i_2(1),\ldots,\mu^i_K(1),\mu^i_1(2),\ldots,\mu^i_K(2),\ldots,\mu^i_1(M),\ldots,\mu^i_K(M)]^T,
\end{equation}
where ${\bm x}^i_k=[x^i_k(1),x^i_k(2),\ldots,x^i_k(M)]^T$ represents the $M\times1$ observation vector at the $k$-th antenna  at the $i$-th observation as in \eqref{eqn:Xn}, and ${\bm \mu}^i_k=[\mu^i_k(1),\mu^i_k(2),\ldots,\mu^i_k(M)]^T$ is the corresponding noise vector at the $k$-th antenna at the $i$-th observation as in \eqref{eqn:Nn}.
Then, the new channel matrix $\bf{H}$ becomes an $MK \times (M+L-1$) matrix:
\begin{equation} \label{eqn:newChannH}
              {\bf H}=\left[ \begin{array}{ccccccccc}
                      h_1(L-1)  &   h_1(L-2)  &   \ldots   &   h_1(0)\\
                      \vdots    &                          &    \vdots\\
                      h_K(L-1)  &   h_K(L-2)  &   \ldots   &   h_K(0)\\
                        &            h_1(L-1) &   h_1(L-2) &   \ldots   & h_1(0)\\
                        &             \vdots                      &       \vdots\\
                        &             h_K(L-1)&   h_K(L-2) &   \ldots    & h_K(0)\\
                        &                    &    \ddots  &    \ddots  \\
                        &              &            &             &           &    h_1(L-1)& h_1(L-2)& \ldots & h_1(0)\\
                        &              &            &             &           &          \vdots    &    \vdots\\
                        &              &            &             &           &    h_K(L-1)& h_K(L-2)& \ldots & h_K(0)
                      \end{array} \right].
\end{equation}
Here, $h_{k}(i)$, for $i=0, \ldots, L-1$, denotes
the $i$-th channel tap observed at $k$-th antenna. To
ensure that $\bf{H}$ is now an over-determined matrix (the order of
row is larger than the order of column), we need to have
\begin{equation} \label{eqn:conK}
    {K>\frac{L+M-1}{M}},
\end{equation}
or, alternatively,
\begin{equation} \label{eqn:conM}
    {M>\frac{L-1}{K-1}}.
\end{equation}
Furthermore, we assume the noise samples come form different antennas are independent with zero mean and ${\cal E}({\bm \mu_i}{\bm \mu^{H}_i})=\sigma^2{\bf I}_{MK}$. Then
we can exactly ensure that the system mode under multiple antennas satisfies the over-determined condition
specified in \cite{Wax85}. For ease of presentation, we define $p=MK$ and $q=L+M-1$ in \eqref{eqn:H1mat}.

As mentioned earlier, the second approach to construct the over-determined channel model is for the secondary user to over-sample the received signals. Suppose that the over-sampling factor is given by $K$. That is, the received baseband signal is sampled $K$ times in one symbol. Then a similar system model as in \eqref{eqn:newXn},  \eqref{eqn:newNn} and \eqref{eqn:newChannH} can be obtained, except that ${\bm x}_i$ and ${\bm \mu}_i$ should be replaced with
\begin{equation} \label{eqn:OvernewXn}
    {{\bm x}_i}={[x(iMK-MK+1),x(iMK-MK+2),\ldots,x(iMK)]^T},
\end{equation}
\begin{equation} \label{eqn:OvernewNn}
    {{\bm \mu}_i}={[\mu(iMK-MK+1),\mu(iMK-MK+2),\ldots,\mu(iMK)]^T},
\end{equation}
and $h_{k}(i)$, for $i=0, \ldots, L-1$, becomes the $k$-th
over-sampling point of the $i$-th channel tap. It can be verified
that $h_{k}(i)$'s are different for different $k$ \cite{Tse2005}.
The major difference between the over-sampling approach and the
multiple-antenna approach is that the over-sampled noise samples in
\eqref{eqn:OvernewNn} are correlated, which contradicts the primary
assumption of independent noise samples. Nevertheless, the
pre-whiting technique can be used to whiten the correlated noises
based on the known correlation matrix. The details can be referred
to Appendix~\ref{prof_Whiten}.

Before leaving this section.
it is noted that, though the proposed over-determined model is based on the assumption that there is only one primary user in the cognitive network, it is also applicable the scenario where there exist multiple primary users. An alternative approach to construct the over-determined model in the presence of multiple  primary users is to use
the cooperative sensing technique as in \cite{Zayen10} by using multiple detectors.

\section{Simplified Information Theoretic Criteria Sensing Algorithm and Performance analysis}
Since the binary hypothesis test in the spectrum sensing is equivalent to the special case of source number estimation problem,
the information theoretic criteria method can be directly applied to conduct spectrum sensing as firstly proposed in \cite{Zayen09, Zayen08, Majed07, Zayen10}. The basic idea is when the primary user is absent, the received signal ${\bm x}_i$ is only the white noise samples. Therefore, the estimated number of source signals via information theoretic criteria (AIC or MDL) should be zero.
Otherwise, when the primary user is present, the number of source signals must be larger than zero.
Hence, by comparing the estimated number of source signals with zero, the presence of the primary user can be detected.
It is noted that the estimation of the number of source by using \eqref{eqn:AICestimate} and \eqref{eqn:MDLestimate} needs very
little prior information about the primary user. In particular, it
does not require the knowledge of channel state information,
synchronization, nor pilot design and modulation strategy. Moreover
it does not need the estimation of noise power. Hence we argue that
information theoretic criteria method is a blind spectrum sensing similar to \cite{Zeng09,zeng08sp,RuiZhang10com}, and it is robust and suitable for the
practical applications.

However, it is known that signal detection is much easier than signal estimation. Therefore, using the estimation method to conduct the detection as in \cite{Zayen09, Zayen08, Majed07, Zayen10} may lead to unnecessary computational complexity overhead.
In the mean time, it makes it difficult to carry out analytical study on the detection performance. In this section,
we propose a simplified ITC algorithm to conduct the spectrum sensing. It can significantly reduce the computational complexity while having almost no performance loss as will be illustrated in Section V.
It also enables a more tractable analytical study on the detection performance.

\subsection{Simplified ITC sensing algorithm}
Before presenting the simplified ITC sensing algorithm in detail, we have the following lemma.

\textbf{Lemma 1}: If there is one value $\hat k (>0)$ which minimizes the AIC metric in \eqref{eqn:AICsolve}  (MDL metric in \eqref{eqn:MDLsolve}), then $\rm {AIC}(0)>\rm {AIC}(1)$ ($\rm {MDL}(0)>\rm {MDL}(1)$) with high probability.

\begin{proof}
Please refer to Appendix~\ref{prof_lemma1}
\end{proof}

The outline of the proposed simplified sensing algorithm is as follows.

\vspace{0.4cm}
\hrule
\hrule
\vspace{0.2cm} \textbf{Algorithm 1: SITC sensing algorithm}
\vspace{0.2cm} \hrule \vspace{0.3cm}

~~~Step 1. Compute the sampled covariance matrix of received
signals, i.e.,
${{\bf R}_x}={\frac{1}{N}\sum^N_{i=1}{{{\bm x}_i}}{{\bm x}_i}^{\dagger}},$
where ${\bm x}_i$'s are received vectors as described in
\eqref{eqn:Xn} or \eqref{eqn:OvernewXn} and $N$ denotes the number of the observations.

~~~Step 2. Obtain the eigenvalues of ${\bf R}_x$ through eigenvalue
decomposition technique, and denote them as $\{l_1,l_2,\ldots,l_p\}$
with $l_1\geq l_2,\ldots,\geq l_p$.

~~~Step 3. Calculate the decision values $\rm {AIC}(0)$ and $\rm {AIC}(1)$ ($\rm {MDL}(0)$ and $\rm {MDL}(1)$) according to \eqref{eqn:AICsolve}(\eqref{eqn:MDLsolve}). Then the detection decision metric is
\begin{equation} \label{eqn:SITC1}
    {\cal T}_{\rm {SITC-AIC}}({\bf L}_x) :\rm {AIC}(0) \mathop \gtrless^{{\cal H}_1}_ {{\cal H}_0}
    \rm {AIC}(1)  .
\end{equation}
if AIC is adopted, or
\begin{equation} \label{eqn:SITC1}
    {\cal T}_{\rm {SITC-MDL}}({\bf L}_x) :\rm {MDL}(0) \mathop \gtrless^{{\cal H}_1}_ {{\cal H}_0}
    \rm {MDL}(1).
\end{equation}
if MDL is adopted, where ${\bf L}_x$ denote the set of eigenvalues $\{l_i,i=1,2,\ldots,p\}$
\vspace{0.2cm} \hrule \vspace{0.4cm}
Note that in the OITC sensing algorithm \cite{Zayen09}, one needs to find the exact value of $\hat k$ from $0$ to $p-1$ to minimize the AIC in \eqref{eqn:AICsolve} or MDL in \eqref{eqn:MDLsolve}. In the proposed SITC algorithm, only two decision values at $k=0$ and $1$ should be computed and compared. Thus, the computational complexity is significantly reduced.
In the next subsection, based on the proposed SITC algorithm, we present the analytical results on the detection performance. Since from the Lemma 1, the SITC algorithm almost obtains the same performance as OITC algorithm, we claim that our analytical results are also applicable for evaluating the performance of OITC algorithm.
\subsection{Performance Analysis}
Since spectrum sensing is actually a binary hypothesis test,
the performance we focus on is the probability of detection $P_d$ (the probability for identifying
the signal when the primary user is present) and the probability of false alarm $P_f$
(the probability for identifying the signal when the primary user is absent).
As no threshold value is involved in the ITC sensing algorithm, $P_d$ is not directly related with $P_f$.
The two probabilities are presented separately.
For ease of presentation, we shall take the AIC criterion for example to illustrate the analysis throughout this section.
The extension to MDL criterion is straightforward if not mentioned otherwise.

\subsubsection{Probability of false alarm}
According to the sensing steps in Algorithm 1, the false alarm occurs when
$\rm {AIC}(0)$ is larger than $\rm {AIC}(1)$ at hypothesis ${\cal H}_0$.
The probability of false alarm can be expressed as
\begin{equation} \label{eqn:PfAIC}
      P_{f-AIC}  = {\rm Pr} \big( {\rm {AIC}}(0) > {\rm {AIC}}(1)| {\cal H}_0 \big).
\end{equation}
Since the primary user is absent, the received signal ${\bm
x}_i$ only contains the noises. The sampled covariance matrix ${\bf
R}_x$ in Algorithm 1 thus turns to ${\bf R}_{\mu}$ defined as
\begin{equation} \label{eqn:Rmu}
    {{\bf R}_\mu}={\frac{1}{N}\sum^N_{i=1}{{{\bm \mu}_i}}{{\bm \mu}_i}^{\dagger}}.
\end{equation}
Hence, the eigenvalues in \eqref{eqn:AICsolve}
become the eigenvalues of the sampled noise covariance matrix
${\bf R}_\mu$ in \eqref{eqn:Rmu}, which is a Wishart random matrix \cite{Johnstone01}.
By applying the recent advances on the
eigenvalue distribution for Wishart matrices, a closed-form expression for the probability
of false alarm can be obtained.

\textbf{Proposition 1}:
The probability of false alarm of the proposed spectrum sensing algorihtm can be approximated as:
\begin{eqnarray} \label{eqn:PfOfAIC}
    {P_f} \approx {F_2}{\left(\frac{pN-(\sqrt{N}+\sqrt{p})^2}{(\sqrt{N}+\sqrt{p})(\frac{1}{\sqrt{N}}+\frac{1}{\sqrt{p}})^{\frac{1}{3}}}\right)} -{F_2}{\left(\frac{(p-\alpha_1)N-(\sqrt{N}+\sqrt{p})^2}{(\sqrt{N}+\sqrt{p})(\frac{1}{\sqrt{N}}+\frac{1}{\sqrt{p}})^{\frac{1}{3}}}\right)}\nonumber\\
    +{F_2}{\left(\frac{(p-\alpha_2)N-(\sqrt{N}+\sqrt{p})^2}{(\sqrt{N}+\sqrt{p})(\frac{1}{\sqrt{N}}+\frac{1}{\sqrt{p}})^{\frac{1}{3}}}\right)}
    -{F_2}{\left(\frac{-(\sqrt{N}+\sqrt{p})^2}{(\sqrt{N}+\sqrt{p})(\frac{1}{\sqrt{N}}+\frac{1}{\sqrt{p}})^{\frac{1}{3}}}\right)},
\end{eqnarray}
where $F_2(\cdot)$ is the cumulative distribution function (CDF) of
Tracy-Widom distribution of order two \cite{Johnstone01}, $\alpha_1$
and $\alpha_2$ with $\alpha_1<\alpha_2$ are the two real roots of the function in
\eqref{eqn:Append-A-funfA} if AIC is applied, or
\eqref{eqn:Append-A-funfAMDL} if MDL is applied.

\begin{proof}
Recall the definition in \eqref{eqn:PfAIC},
to compute the probability of false alarm
is to compute the probability
\begin{equation} \label{eqn:Append-A-pfAIC}
    {P_{f-AIC}}= {{\rm Pr}({\rm AIC}(0)-{\rm AIC}(1)>0|{\cal H}_0)}.
\end{equation}
According to the cost function of AIC defined in \eqref{eqn:AICsolve}, we have
\begin{equation} \label{eqn:Append-A-AIC0-1}
    {{\rm AIC}(0)-{\rm AIC}(1)}={-2\log{\left[\frac{\prod^p_{i=1}{l_i^{1/p}}}{{\frac{1}{p}}{\sum^p_{i=1}{l_i}}}\right]^{pN}}}
    {+2\log{\left[\frac{\prod^p_{i=2}{l_i^{1/{p-1}}}}{{\frac{1}{p-1}}{\sum^p_{i=2}{l_i}}}\right ]^{(p-1)N}}-(4p-2)}.\nonumber
\end{equation}
Then we can rewrite \eqref{eqn:Append-A-pfAIC} as
\begin{equation} \label{eqn:Append-A-pfAIC2}
    {P_{f-AIC}} = {\rm Pr} \left( {\log}\left[\frac{({\frac{1}{p}}{\sum^p_{i=1}}{l_i})^p}
    {({{\frac{1}{p-1}}{\sum^p_{i=2}{l_i}}})^{p-1}l_1} \right]  > \frac{4p-2}{2N} \bigg|{\cal H}_0\right ).
\end{equation}
Note here, the sum of eigenvalues of sampled covariance matrix,
$\frac{1}{p}\sum^p_{i=1}{l_i}$,
is equivalent to ${\frac{1}{pN}}{\rm Tr}\left({\sum^N_{i=1}{{{\bm x}_i}{{{\bm x}_i}}^{\dagger}}}\right)$.
At hypothesis $H_0$, where the received vector involves only the noise samples,
${\frac{1}{pN}}{\rm Tr}\left({\sum^N_{i=1}{{{\bm x}_i}{{{\bm x}_i}}^{\dagger}}}\right)$
is the un-biased estimation of the covariance of the white noise.
Therefore, when $N$ is sufficiently large, we have
\begin{equation} \label{eqn:Append-A-estnoise}
    {\frac{1}{p}\sum^p_{i=1}{l_i}}\approx {\sigma^2}.
\end{equation}
Substituting \eqref{eqn:Append-A-estnoise} into \eqref{eqn:Append-A-pfAIC2} yields:
\begin{equation} \label{eqn:Append-A-pfAIC3}
    {P_{f-AIC}} \approx {\rm Pr}{\left[ {\frac{(\sigma^2)^p}{(\frac{p}{p-1}\sigma^2-\frac{l_1}{p-1})^{p-1}l_1}}>\exp{\left( \frac{2p-1}{N}\right)} \bigg|{\cal H}_0\right]}.
\end{equation}
From \eqref{eqn:Append-A-pfAIC3}, it is seen that the probability of false alarm
is only dependent on the largest eigenvalue of the noise sampled covariance matrix ${\bf R}_\mu$.
Since ${\bf R}_\mu$ is actually a Wishart random matrix ,
its the largest eigenvalue $l_1$ satisfies the Tracy-Widom distribution of order two \cite{Johnstone01}.
To apply this result, we rewrite \eqref{eqn:Append-A-pfAIC3} as
\begin{eqnarray} \label{eqn:Append-A-pfAIC4}
    {P_{f-AIC}}\approx {\rm Pr} \left[{\frac{l_1}{\sigma^2}}{\left(p-\frac{l_1}{\sigma^2}\right)^{p-1}}<  \frac{(p-1)^{p-1}}{\exp \left(\frac{2p-1}{N} \right)} \bigg|{\cal H}_0\right]
\nonumber \\
    = {\rm Pr}\left[x^p-px^{p-1}+ \frac{(p-1)^{p-1}}{\exp\left(\frac{2p-1}{N}\right)} >0 \bigg|{\cal H}_0\right],
\end{eqnarray}
where $x \triangleq p-\frac{l_1}{\sigma^2}$.

Define a function
\begin{equation}\label{eqn:Append-A-funfA}
    f(x) \triangleq x^p-px^{p-1}+\frac{(p-1)^{p-1}}{\exp\big(\frac{2p-1}{N}\big)}.
\end{equation}
We next find the real roots of this function.

Taking the differentiation of $f(x)$ and equating it to zero, we obtain
\begin{equation}\label{eqn:Append-A-diffefA}
    {\frac{df(x)}{dx}}=px^{p-1}-p(p-1)x^{p-2}=px^{p-2}[x-(p-1)]=0.\nonumber
\end{equation}
Clearly, $f(x)$ has two stationary points, which are $x=p-1$ and
$x=0$. In the following, two scenarios with p being even or odd
are considered respectively. When $p$ is even, it is easily found
that the function $f(x)$ monotonically decreases over $(-\infty,
p-1)$ and monotonically increases over $(p-1,\infty)$.
Simultaneously, we can verify that
\begin{equation}\label{eqn:Append-A-fAmin}
    f(p-1)=(p-1)^p-p(p-1)^{p-1}+\frac{(p-1)^{p-1}}{\exp\big(\frac{2p-1}{N}\big)}<0.
\end{equation}
and
\begin{equation}\label{eqn:Append-A-fA0p}
    f(0)=f(p)=\frac{(p-1)^{p-1}}{\exp\big(\frac{2p-1}{N}\big)}>0.
\end{equation}
So there must be two real real roots within $(0,p)$ and around $p-1$ for function $f(x)$.
Let $\alpha_1$ and $\alpha_2$, with $\alpha_1<\alpha_2$, denote the two real roots,
then \eqref{eqn:Append-A-pfAIC4}
is converted into an equivalent form:
\begin{equation}\label{eqn:Append-A-pfAIC66}
    {P_{f-AIC}}\approx  {\rm Pr}\left[x<\alpha_1 |H_0\right]+{\rm Pr}\left[\alpha_2<x |{\cal H}_0\right].
\end{equation}
When $p$ is odd, we can find $f(x)$ decreases monotonically over $(0,p-1)$, while it is the monotonic
increasing function over both $(-\infty,0)$ and $(p-1,\infty)$.
According to the fact that $f(-\infty)<0$, $f(0)>0$, $f(p-1)<0$ and $f(p)>0$, we conclude that
$f(x)$ have three real roots, which are denoted as $\alpha_0$, $\alpha_1$ and $\alpha_2$, with
$\alpha_0<0$ and $0<\alpha_1<\alpha_2$, respectively.
Then \eqref{eqn:Append-A-pfAIC4}
can be rewritten as:
\begin{equation}\label{eqn:Append-A-pfAIC77}
    {P_{f-AIC}}\approx {\rm Pr}\left[\alpha_0<x<\alpha_1 |H_0\right]+{\rm Pr}\left[\alpha_2<x |{\cal H}_0\right].
\end{equation}
However, it is noted that as $N$ is large enough, the largest
eigenvalue of the sampled noise covariance matrix, $l_1$, is just
slightly larger than the true covariance of noise $\sigma^2$. Hence,
from the definition, $x$ can be reasonably limited in $(0,p)$.
Therefore, both the probability of \eqref{eqn:Append-A-pfAIC66} and
\eqref{eqn:Append-A-pfAIC77} can be summarized as the following form
\begin{equation}
    {P_{f-AIC}}\approx  {\rm Pr}\left[0<x<\alpha_1 |H_0\right]+{\rm Pr}\left[\alpha_2<x<p
    |{\cal H}_0\right]. \nonumber
\end{equation}
In other words,
\begin{equation}\label{eqn:Append-A-pfAIC7}
    {P_{f-AIC}} \approx  {\rm Pr}\left[p-\alpha_1<\frac{l_1}{\sigma^2}<p \bigg|{\cal H}_0\right]+\left[0<\frac{l_1}{\sigma^2}<p-\alpha_2 \bigg|{\cal H}_0\right].\nonumber
\end{equation}
Applying the distribution of the largest eigenvalue for Wishart
matrix in random matrix theory \cite{Johnstone01}, the variable
$N\frac{l_1}{\sigma^2}$ satisfies the distribution of Tracy-widom of
order two, i.e.,
\begin{equation}\label{eqn:Append-A-eigdis}
    \frac{N\frac{l_1}{\sigma^2}-(\sqrt{N}+\sqrt{p})^2}{(\sqrt{N}+\sqrt{p})\big(\frac{1}{\sqrt{N}}+\frac{1}{\sqrt{p}}\big)^{\frac{1}{3}}}\backsim W_2\backsim F_2.\nonumber
\end{equation}
Here, $W_2$ and $F_2$ denote the probability density function (PDF)
and cumulative density function (CDF) for distribution of
Tracy-widom of order two respectively. Therefore, the probability of
false alarm of AIC can be concluded as \eqref{eqn:PfOfAIC}.

Similar with the above derivation, when the MDL criterion is applied, we only need to modify the step in \eqref{eqn:Append-A-pfAIC4} as
\begin{equation}\label{eqn:Append-A-pfMDL1}
    {P_{f-MDL}} \approx {\rm Pr}\left[x^p-px^{p-1}+ \frac{(p-1)^{p-1}}{\exp\left(\frac{(p-0.5)\log{N}}{N}\right)} >0 \bigg|{\cal H}_0\right].\nonumber
\end{equation}
and redefine the function $f(x)$ in \eqref{eqn:Append-A-funfA} as
\begin{equation}\label{eqn:Append-A-funfAMDL}
    f(x)\triangleq x^p-px^{p-1}+\frac{(p-1)^{p-1}}{\exp\big(\frac{(p-0.5)\log{N}}{N}\big)}.
\end{equation}
\end{proof}

From Proposition 1, it is found that the probability of false alarm
is independent with noise covariances $\sigma^2$. Therefore, the
proposed SITC sensing algorithm is
robust in practical applications. It is also noted that $P_f$ depends on the product of $M$ and $K$, i.e., $p=MK$, rather than the individual values of $M$ and $K$.
\subsubsection{Probability of detection}
When the primary user is present, the event of detection also occurs when $\rm {AIC}(0) > \rm {AIC}(1)$.
The probability of detection is thus described as
\begin{equation} \label{eqn:PdAIC}
    P_{d-AIC}  = {\rm Pr} \big( {\rm {AIC}}(0) > {\rm {AIC}}(1) | {\cal H}_1 \big).
\end{equation}
Since at ${\cal H}_1$, the received vector ${\bm x}_i$ involves the signals transmitted by the primary user, the sampled covariance
matrix ${\bf R}_x$ can be written as
\begin{equation} \label{eqn:Rxn}
    {{\bf R}_x}={\frac{1}{N}} \sum^N_{i=1} {({\bf H}{\bm s}_i+{\bm \mu}_i)({\bf H}{\bm s}_i+{\bm
    \mu}_i)^{\dagger}}.
\end{equation}
Note that ${\bf R}_x$ is no longer a Wishart matrix. The exact distribution of its eigenvalues
is unknown and difficult to find, and hence so is the $P_d$.
In the following, we resort to deriving a closed-form expression
for the conditional probability of detection given the channel matrix $\bf H$.
The average probability of detection can then be obtained using a hybrid analytical-simulation
approach.

\textbf{Proposition 2}: Let ${\bf R}_s$ denote the covariance matrix of ${\bm s}_i$
given in \eqref{eqn:Sn} and $\{\delta_1, \delta_2, \ldots, \delta_p\}$
be the eigenvalues of matrix ${\bf H}{\bf R}_s{\bf H}^\dagger$ (with $\delta_1 \geqslant \delta_2 \geqslant \ldots \geqslant \delta_p$).
Then there exists a value of $\rho$, for $\delta_p \leqslant \rho \leqslant \delta_1$, such that
the probability of detection given $\bf H$ can be approximated as
$P_{d|H}\approx Q(\rho)$, where the function $Q(\cdot)$ is
\begin{eqnarray} \label{eqn:Qfun}
    {Q(\delta)}={F_2}{\left(\frac{pN-(\sqrt{N}+\sqrt{p})^2}{(\sqrt{N}+\sqrt{p})(\frac{1}{\sqrt{N}}+\frac{1}{\sqrt{p}})^{\frac{1}{3}}}\right)}
    -{F_2}{\left(\frac{(\frac{(p-\pi_1)\epsilon-\delta}{\sigma_2})N-(\sqrt{N}+\sqrt{p})^2}{(\sqrt{N}+\sqrt{p})(\frac{1}{\sqrt{N}}+\frac{1}{\sqrt{p}})^{\frac{1}{3}}}\right)}\nonumber \\
    +{F_2}{\left(\frac{(\frac{(p-\pi_2)\epsilon-\delta}{\sigma_2})N-(\sqrt{N}+\sqrt{p})^2}{(\sqrt{N}+\sqrt{p})(\frac{1}{\sqrt{N}}+\frac{1}{\sqrt{p}})^{\frac{1}{3}}}\right)}
    -{F_2}{\left(\frac{-(\sqrt{N}+\sqrt{p})^2}{(\sqrt{N}+\sqrt{p})(\frac{1}{\sqrt{N}}+\frac{1}{\sqrt{p}})^{\frac{1}{3}}}\right)},
\end{eqnarray}
where $\epsilon={\frac{1}{p}{\rm Tr}(\bf{H}\bf{R}_s\bf{H}^\dagger)+\sigma^2}$ and $\pi_1$, $\pi_2$ (with $\pi_1<\pi_2$)
denote the two roots of the function \eqref{eqn:Append-c-jfun} for AIC or
\eqref{eqn:Append-c-jfun2} for MDL.
Furthermore, upper and lower bounds can be
obtained as $Q(\delta_p)\leqslant P_{d|H} \leqslant Q(\delta_1)$.

\begin{proof}
Please refer to Appendix~\ref{prof_pro3}.
\end{proof}

From Proposition 2, we find that $P_d$ is not only related to $N$
and $p$, but also depends on $\frac {\epsilon}{\sigma^2}$, which is
the ratio of the signal strength of  primary user to the noise
variance. The exact value of $\rho\in [\delta_p,\delta_1]$  in
Proposition 2 is difficult to determine in an analytical way, since
it is related to both the channel response $\bf H$ and the
covariance matrix of source signal ${\bf R}_s$. In practice, we can
simply choose $\rho_{AIC}=\frac{1}{2}(\delta_p+\delta_1)$ and
$\rho_{MDL}=\frac{3}{4}(\delta_p+\delta_1)$. It will be demonstrated
later in Section VI that the analytical $P_{d|H}$ based on this
choice of $\rho$ can approximate the Monte Carlo results very well
in most of cases.

\section{Generalized information theoretic criteria sensing algorithm}
As mentioned in the previous section, the probability of detection of and probability
of false alarm of the proposed simplified ITC sensing algorithm are not directly related to each other as the algorithm does not involve any threshold (same for the original ITC algorithm in \cite{Zayen09}). According to the analytical results given in \eqref{eqn:PfOfAIC} and
\eqref{eqn:Qfun}, to satisfy different system requirements,
a proper set of values for the parameters
$M$, $K$ and $N$ in model \eqref{eqn:H1mat}
should be chosen, which is inconvenient for practical application.
In this section, based on the analytical discussion in Section IV, we propose a generalized information theoretic criteria sensing algorithm which can provide a flexible tradeoff between $P_d$ and $P_f$ according to different system design requirements.

From the expression given in \eqref{eqn:Append-A-pfAIC2} and \eqref{eqn:Append-c-pmfull}, we found that the sensing decision for SITC algorithm is actually based on the decision variable ${\log}\left[\frac{({\frac{1}{p}}{\sum^p_{i=1}}{l_i})^p}{({{\frac{1}{p-1}}{\sum^p_{i=2}{l_i}}})^{p-1}l_1} \right]$. Thus, we generalize the decision rule as
\begin{equation} \label{eqn:GITC1}
    {\cal T}_{\rm {GITC}}({\bf L}_x) : \frac{({\frac{1}{p}}{\sum^p_{i=1}}{l_i})^p}
    {({{\frac{1}{p-1}}{\sum^p_{i=2}{l_i}}})^{p-1}l_1}
   \mathop \gtrless^{{\cal H}_1}_ {{\cal H}_0}\gamma,
\end{equation}
where $\gamma$ is a pre-set threshold.
It is seen that if we set $\gamma=\exp{(\frac{2p-1}{N})}$, the decision rule given in \eqref{eqn:GITC1} turns into the AIC based SITC sensing algorithm presented in Algorithm 1. If we fix
$\gamma=\exp{(\frac{(p-0.5)\log{N}}{N})}$, the algorithm becomes the MDL-based SITC sensing algorithm. Furthermore, it is easy to find that the analytical results obtained in Section IV are applicable for the GITC sensing algorithm.
The only change  that needs to be made is to replace $\alpha_1$
and $\alpha_2$ in \eqref{eqn:PfOfAIC} (or $\pi_1$ and $\pi_2$ in \eqref{eqn:Qfun})
by two real roots generated by the following equation.
\begin{equation}\label{eqn:GITC2}
    f(x)\triangleq x^p-px^{p-1}+\frac{(p-1)^{p-1}}{\gamma}.
\end{equation}

Thus, the outline of the proposed GITC sensing algorithm can be summarized as follows.

\vspace{0.4cm}
\hrule
\hrule
\vspace{0.2cm} \textbf{Algorithm 2: GITC sensing algorithm} \vspace{0.2cm} \hrule \vspace{0.3cm}

~~~Step 1 and Step 2: the same as Algorithm 1 in Section IV.

~~~Step 3: According to the system requirement on $P_f$, choose a
proper threshold $\gamma$ based on \eqref{eqn:PfOfAIC} and \eqref{eqn:GITC2}.

~~~Step 4:  Conduct the
decision based on \eqref{eqn:GITC1}.
\vspace{0.2cm} \hrule \vspace{0.4cm}

According to the decision variable presented in \eqref{eqn:GITC1}, we find that
the proposed GITC sensing algorithm is actually also an eigenvalue-based method similar to \cite{Zeng09,zeng08sp,RuiZhang10com}. The advantage of the proposed GITC over the algorithms in \cite{zeng08sp,RuiZhang10com} is that
it is able to analytically obtain the explicit decision threshold
$\gamma$ according to the system requirement on $P_f$ before the actual deployment.

\section{Simulation results and discussions}
In this section, we present some numerical examples to demonstrate
the effectiveness of the proposed sensing schemes and to confirm the
theoretical analysis.
\subsection{Comparison between simulation and analytical results for both SITC and OITC}
In our first set of examples, we compare the simulation results with analytical results given in Proposition 1 and Proposition 2.
The comparison between SITC and OITC are also presented.
In the simulation,
the channel taps are random numbers with zero-mean complex Gaussian
distribution. All the results are averaged over 1000 Monte Carlo
realizations. For each realization, random channel, random noise and
random BPSK modulated inputs are generated. We define the SNR as
the ratio of the average received signal power to the average noise
power
\begin{equation} \label{eqn:SNR}
    {SNR}=\frac {{\cal E}[\|{\bm x}_i-{\bm \mu}_i\|^2]}{{\cal E}[\|{\bm \mu}_i\|^2]}.
\end{equation}

The comparison of simulation and analytical results for $P_f$ is
demonstrated in TABLE~\ref{TABLE:ChangeMKL} and TABLE~\ref{TABLE:ChangeN}. According to Proposition 1, $P_f$ is
independent with the noise variance, thus remains constant over
different SNR. Hence we average multiple values over different SNR
as the simulated $P_f$ and compare it with the analytical $P_f$.
From TABLE~\ref{TABLE:ChangeMKL}, we first observe that SITC and OITC perform almost the same.
It is also seen
that, for AIC, the analytical results are slightly larger than the
simulation results especially when $p=MK$ is small. Nevertheless, the analytical
approximation is accurate enough to evaluate the performance of the
proposed sensing scheme. It is also found that $P_{f-AIC}$ gradually
decreases as $p=MK$ increases while the $P_{f-MDL}$ remains zero in both
simulation and analytical results. We conclude that the MDL method
has excellent false alarm performance.
From TABLE~\ref{TABLE:ChangeN}, we find that the probability of false alarm increases very
slowly as $N$ increases. In fact, our simulation shows that
$P_{f-AIC}$ is still below $0.1$ even when $N=10^{15}$ at $M=5$ and $K=4$.

\begin{figure}[b]
\begin{centering}
\includegraphics[scale=0.68]{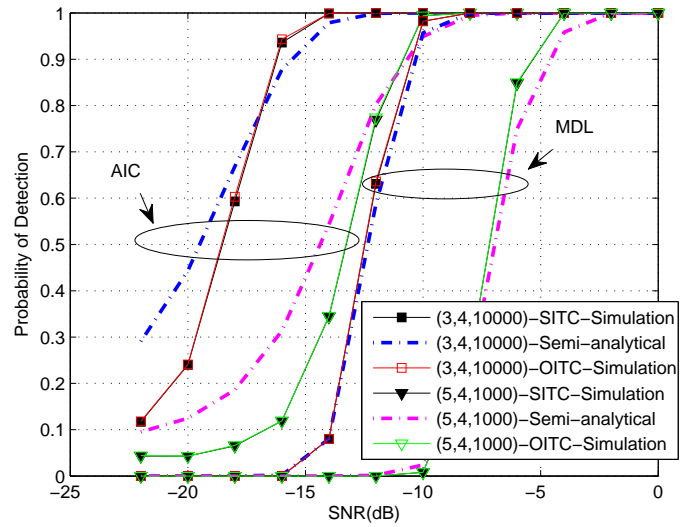}
\vspace{-0.1cm}
\caption{Simulation and theoretic results about probability of detection at different $(M,K,N)$ for both SITC and OITC.} \label{fig:Theoretic_Pd}
\end{centering}
\vspace{-0.3cm}
\end{figure}

\begin{figure}[tbhp]
\begin{centering}
\includegraphics[scale=0.68]{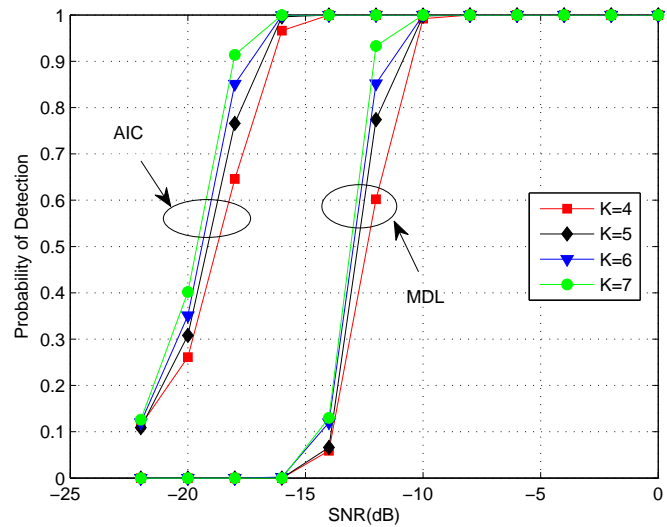}
\vspace{-0.1cm}
\caption{Probability of detection for different $K$ at $M=5$ and $N=10000$.} \label{fig:ChangeK}
\end{centering}
\vspace{-0.3cm}
\end{figure}

\begin{figure}[tbhp]
\begin{centering}
\includegraphics[scale=0.68]{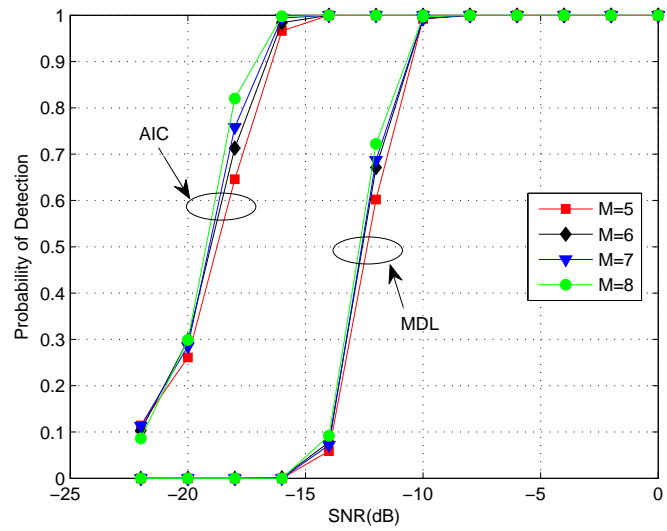}
\vspace{-0.1cm}
\caption{Probability of detection for different $M$ at $K=4$ and $N=10000$.} \label{fig:ChangeM}
\end{centering}
\vspace{-0.3cm}
\end{figure}

\begin{figure}[tbhp]
\begin{centering}
\includegraphics[scale=0.68]{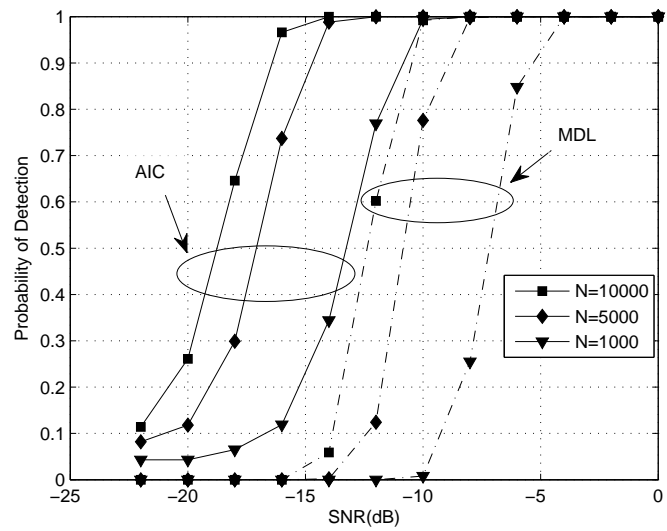}
\vspace{-0.1cm}
\caption{Probability of detection for different $N$ at $M=5$ and $K=4$.} \label{fig:ChangeN}
\end{centering}
\vspace{-0.3cm}
\end{figure}

Figs.~\ref{fig:Theoretic_Pd}-\ref{fig:ChangeN} show $P_d$ at
different system parameters. In Fig.~\ref{fig:Theoretic_Pd},
we first compare the
detection performance obtained by simulation between SITC and OITC.
It is seen that the proposed SITC sensing algorithm do not lead to any performance loss
compared to OITC algorithm.
Then, comparing the semi-analytical results obtained from Proposition 2 with the simulation results, one can observe a very good match between them, especially for MDL method.
Thus, Proposition 2 is validated.
Fig.~\ref{fig:ChangeK}, Fig.~\ref{fig:ChangeM} and
Fig.~\ref{fig:ChangeN} present
the simulation results of $P_d$ for variable $K$ (at
$M=5, N=10000$), $M$ (at $K=4, N=10000$) and $N$ (at $K=4, M=5$),
respectively. It is found that the
performance is improved as any of these parameters increases.

\subsection{Comparison between SITC and other sensing algorithms}
\begin{figure}[tbhp]
\begin{centering}
\includegraphics[scale=0.68]{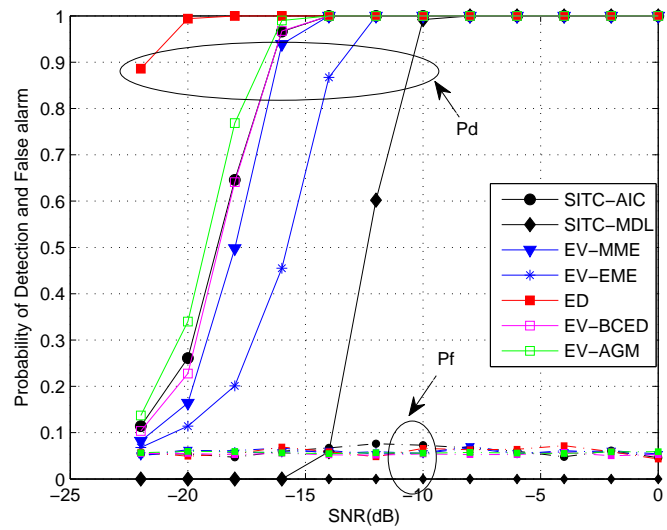}
\vspace{-0.1cm} \caption{Comparison with the eigenvalue-based
methods and the energy detection method at $M=5$, $K=4$ and
$N=10000$.} \label{fig:ComparisonWithLiang}
\end{centering}
\vspace{-0.3cm}
\end{figure}

Thus far, a few efficient sensing algorithms have been proposed in
the literature, with each requirng distinct prior information. In
this subsection, for fair comparison, we only choose the
eigenvalue-based methods proposed in
\cite{Zeng09,zeng08sp,RuiZhang10com} and the energy detection method
since they both need little prior information. It should be
mentioned that the proposed SITC-AIC and SITC-MDL algorithms are
equivalent to the GITC algorithm provided in Section V via setting
$\gamma_{\rm AIC}=\exp{(\frac{(2p-1)}{N})}$ and $\gamma_{\rm
MDL}=\exp{(\frac{(p-0.5)\log{N}}{N})}$. Therefore, we omit the
performance comparison with the GITC. In the simulation, we fix the
order of channel $L=10$ as in \cite{Zeng09} and choose $N=10000$,
$K=4$ and $M=5$. Fig.~\ref{fig:ComparisonWithLiang} shows the
comparison with the energy detection (ED) method (with perfect
estimation of noise covariance) and the four eigenvalue-based
methods, namely, the maximum minimum eigenvalue detection (EV-MME)
and energy with minimum eigenvalue detection (EV-EME) \cite{Zeng09},
the blindly combined energy detection (EV-BCED) \cite{zeng08sp}, and
the arithmetic to geometric mean (EV-AGM) \cite{RuiZhang10com}. We
see that, under almost the same $P_f$, energy detection performs the
best, followed by the EV-AGM method, the proposed SITC-AIC method
and EV-BCED method, and then EV-MME and EV-EME methods. Among the
proposed SITC-AIC and four eigenvalue-based methods, the SITC-AIC
almost obtains the same performance with the EV-BCED method and they
both outperform EV-MME and EV-EME while being slightly inferior to
the EV-AGM method. Though the proposed SITC-MDL method performs the
worst in $P_d$, it is the best among all the considered schemes in
terms of $P_f$ performance.

\begin{figure}[tbhp]
\begin{centering}
\includegraphics[scale=0.68]{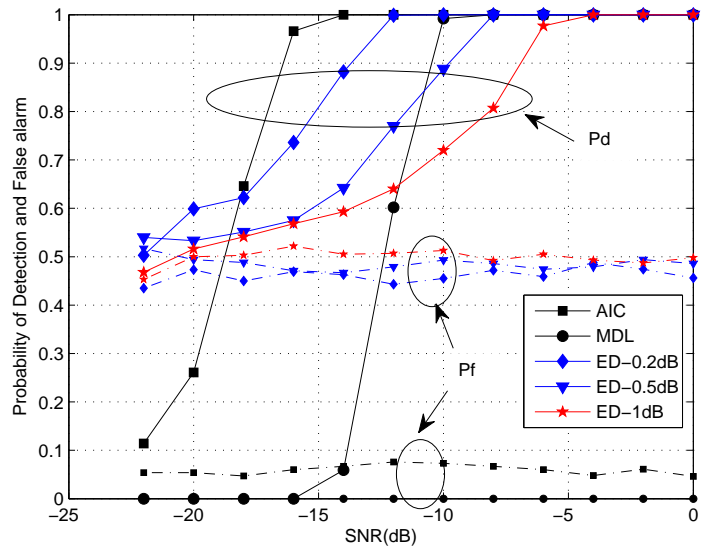}
\vspace{-0.1cm}
\caption{Comparison with energy detection with noise uncertainty at $M=5$, $K=4$ and $N=10000$.} \label{fig:ComparisonWithED}
\end{centering}
\vspace{-0.3cm}
\end{figure}

The comparison with energy detection with noise uncertainty is
presented in Fig.~\ref{fig:ComparisonWithED}, where ``ED-x dB" means
that the noise uncertainty in energy detection is x-dB as defined in
\cite{Zeng09}. It is observed that, although the proposed method performs
worse than the energy detection method with accurate noise
covariance estimation, it significantly outperforms in both $P_d$
and $P_f$ when there exists some noise uncertainty. This clearly
demonstrates the robustness of information theoretic criteria based
blind sensing algorithm.

\subsection{Performance of the GITC algorithm}
\begin{figure}[tbhp]
\begin{centering}
\includegraphics[scale=0.68]{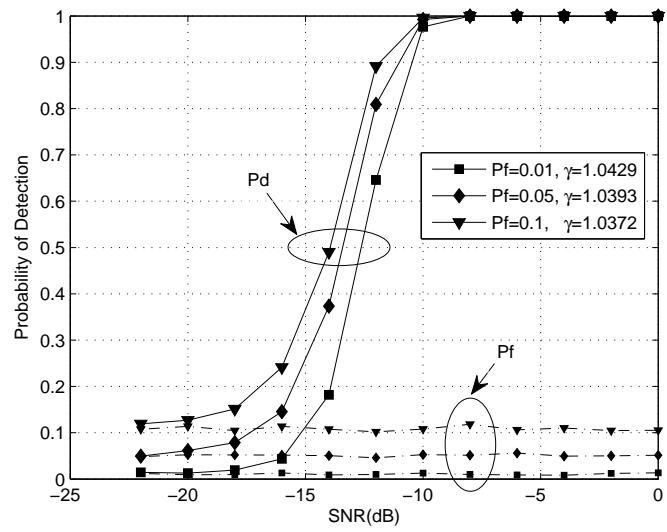}
\vspace{-0.1cm}
\caption{Simulation results for GITC algorithm for different $P_f$ at $M=5$, $K=4$ and $N=1000$.} \label{fig:ThresholdCase}
\end{centering}
\vspace{-0.3cm}
\end{figure}

Results for the GITC sensing algorithm at
different threshold values are demonstrated in
Fig.~\ref{fig:ThresholdCase}.
It is assumed that we should choose proper thresholds to make $P_f=0.1$, $P_f=0.05$ and $P_f=0.01$.
According to the Proposition 1 and the discussion in Section V, we choose three
thresholds $\gamma=1.0372$, $\gamma=1.0393$ and $\gamma=1.0429$
(note that since the analytical results are slightly larger than the simulation results,
the thresholds we choose should make theoretic $P_f$ larger than required $P_f$ by about 0.02).
From the plots, it is found that the $P_f$ requirements are satisfied very well. One can also see that the probability of false
alarm is very sensitive to the threshold. Hence, the GITC
sensing algorithm is flexible for system design with different requirements.

\section{Conclusions}
In this paper we have provided an intensive study on the information theoretic criteria based blind spectrum sensing method. Based on the prior work on the related study, we first proposed the simplified ITC sensing algorithm.
This algorithm significantly reduces the
computational complexity without losing any detection performance compared with the existing ITC based sensing algorithm.
Moreover, it enables a more trackable analytical study on the detection performance.
Thereafter, applying the recent advances
in random matrix theory, we derive closed-form expressions for both the probability of false
alarm and probability of detection which can tightly approximate the actual results in simulation.
We further generalized the SITC sensing algorithm to an eigenvalue based sensing algorithm which strike the balance between the probabilities of detection and
false alarm by involving an adjustable threshold. Simulation results demonstrate that the proposed blind sensing algorithm outperforms the existing eigenvalue-based sensing algorithms in certain scenarios.

\appendices
\section{Whitening the over-sampled noises}
\label{prof_Whiten} At the secondary receiver, the received
continuous signal is usually filtered by a low-pass filter.
Therefore, the noise $\mu(t)$ in \eqref{eqn:H0con} and
\eqref{eqn:H1con} should be correlated. We assume that the white
noise before the filter is ${\hat \mu}(t)$ and the system function
of the low-pass filter is $g(t)$ which is known at the secondary
receiver. In the following, we only consider the real value case,
since in the communication system, the complex value signal is just
the combination of two orthogonal real value signals. As we have
known, $\mu(t)$ can be described by ${\hat \mu}(t)$ and $g(t)$ as
\begin{equation}\label{eqn:Append1-1}\nonumber
     \mu(t)=g(t)\otimes {\hat \mu}(t) =\int^{t_{max}}_0 {g(\ell){\hat \mu}(t-\ell)}{d\ell},
\end{equation}
where $(0,t_{max})$ represents the time span of $g(t)$ and $\otimes$
denotes the convolution operator. Thus, the auto-correlative
function of $\mu(t)$ denoted by $\phi_{\mu}(\tau)$ can be expressed
as
\begin{equation}\label{eqn:Append1-2}\nonumber
     \phi_{\mu}(\tau)=\phi_{g}(\tau)\otimes\phi_{{\hat \mu}}(\tau),
\end{equation}
where $\phi_{g}(\tau)$ and $\phi_{{\hat \mu}}(\tau)$ are the
auto-correlative functions of $g(t)$ and ${\hat \mu}(t)$,
respectively. Note that $\phi_{{\hat \mu}}(\tau)$ should be equal to
$\sigma^2\delta(\tau)$ since ${\hat \mu}(t)$ is white (here the
covariance of ${\hat\mu}(t)$  is assumed to be $\sigma^2$ ).
Therefore, we derive that
\begin{equation}\label{eqn:Append1-3}\nonumber
     \phi_{\mu}(\tau)=\sigma^2\phi_{g}(\tau)=\sigma^2\int^{t_{max}}_0 {g(\ell)g(\tau-\ell)}{d\ell},~~0\leq\tau\leq2t_{max}
\end{equation}
Thus, if the received signal is over-sampled at rate $Kf_s$ where
$f_s$ is the reciprocal of the baseband symbol duration $T_0$ and
$K$ is the over-sampling factor, the covariance matrix of the noise
vector ${\bm \mu}_i$ given in \eqref{eqn:OvernewNn} becomes
\begin{equation}\label{eqn:Append1-4}\nonumber
     {\bf R}_{\mu}=\sigma^2{\bf Q},
\end{equation}
with $\bf Q$ having entries $q_{i,j}=\phi_{g}(|i-j|\frac{T_0}{K})$.
Note that $\bf Q$ is a positive definite symmetric matrix. It can be
decomposed into ${\bf Q}={\tilde {\bf Q}}^2$, where $\tilde {\bf Q}$
is also a positive definite symmetric matrix. Hence, to obtain the
independent noise samples in the over-sampling scheme, we can
pre-whiten the over-sampled noise samples ${\bm \mu}_i$ as
\begin{equation}\label{eqn:Append1-5}\nonumber
     {\tilde {\bm \mu}_i}={\tilde {\bf Q}}^{-1}{\bm \mu}_i.
\end{equation}
Then, the covariance matrix of $\tilde {\bm \mu}_i$ transforms into
\begin{equation}\label{eqn:Append1-6}\nonumber
     {\bf R}_{{\tilde{\mu}}_i}={\tilde {\bf Q}}^{-1}{\bf R}_{\mu}{\tilde {\bf Q}}^{-1}=\sigma^2{\bf I}_p.
\end{equation}
Now, noise samples ${\bm \mu}_i$ are whitened. It is noted that
$\tilde {\bf Q}$ is only related to the low-pass filter and
over-sampling factor $K$ and is independent to the signal and noise.
Therefore, the pre-whitening process can be used blindly.

\section{Proof of Lemma 1}
\label{prof_lemma1} We prove the lemma from two aspects. Firstly,
%
it has been shown in \cite{Xu95, Fishler02} that most of the
estimation errors of AIC and MDL occur tightly around the true
numbers. According to this finding, at hypothesis ${\cal H}_0$ (the
true number of source signal is zero), if there exists ${\hat k}>0$
minimizing \eqref{eqn:AICsolve} or \eqref{eqn:MDLsolve}, then we
have ${\hat k}=1$ with high probability. Hence, Lemma 1 holds for
the case of false alarm.
Next, we prove that Lemma 1 succeeds at hypothesis ${\cal H}_1$.
Since the primary user is present, the eigenvalues $l_i$ of the
sampled covariance matrix are distinct at least for $i=1,2,\ldots,q$
(here $q$ is the true source number). For $i=q,q+1,\ldots,p$, the
eigenvalues are actually the estimation of noise variance
$\sigma^2$. They may be equal to each other when $N$ is enough
large. According to the expression of AIC and MDL, it is found that
the second terms in \eqref{eqn:AICsolve} and \eqref{eqn:MDLsolve}
are monotonically increasing functions of $k$. To make the cost
function in \eqref{eqn:AICsolve} or \eqref{eqn:MDLsolve} minimum at
${\hat k}\in [1,p-1]$, we must have that the first terms in
\eqref{eqn:AICsolve} and \eqref{eqn:MDLsolve} are monotonically
decreasing for $k=0,1,\ldots,\hat k$. We next prove this statement.

We focus on the AIC criterion and the extension to MDL  is
straightforward. Supposing $k'\in[2,{\hat k}]$ and
\begin{equation}\label{eqn:Append2-1}\nonumber
     f_{\rm AIC}(k)={-2\log \left (\frac{\prod^p_{i=k+1}{l^{1/(p-k)}_i}}{{\frac{1}{p-k}}\sum^p_{i=k+1}{l_i}}\right )^{(p-k)N}},
\end{equation}
we have
\begin{equation}\label{eqn:Append2-2}\nonumber
     f_{\rm AIC}(k'-1)-f_{\rm AIC}(k')=2N\log \frac{\left(\frac{1}{p-k'+1}\sum^p_{i=k'}l_i \right)^{p-k'+1}}
     {\left(\frac{1}{p-k'}\sum^p_{i=k'+1}l_i\right)^{p-k'}l_k'}.
\end{equation}
Since
\begin{equation}\label{eqn:Append2-3}\nonumber
\begin{split}
     &\left(\frac{1}{p-k'+1}\sum^p_{i=k'}l_i \right)^{p-k'+1} \\
     &=\left(\frac{1}{p-k'}\frac{p-k'}{p-k'+1}\sum^p_{i=k'+1}l_i+\frac{1}{p-k'+1}l_{k'}\right)^{p-k'+1}\\
     &\geq
     \left[\left(\frac{1}{p-k'}\sum^p_{i=k'+1}l_i\right)^{\frac{p-k'}{p-k'+1}}\right.
     \left.l_k'^{\frac{1}{p-k'+1}}\right]^{p-k'+1}\\
     &=\left(\frac{1}{p-k'}\sum^p_{i=k'+1}l_i\right)^{p-k'}l_k'
\end{split}
\end{equation}
(here, the arithmetic-mean geometric-mean inequality $x_1^{a_1}+x_2^{a_2}\geq x_1^{a_1}x_2^{a_2}$ with $a_1+a_2=1$ is applied),
we conclude that
\begin{equation}\label{eqn:Append2-3}\nonumber
    \frac{\left(\frac{1}{p-k'+1}\sum^p_{i=k'}l_i \right)^{p-k'+1}}
     {\left(\frac{1}{p-k'}\sum^p_{i=k'+1}l_i\right)^{p-k'}l_k'}
     \geq 1.
\end{equation}
It further means
\begin{equation}\label{eqn:Append2-4}\nonumber
   f_{\rm AIC}(k'-1)-f_{\rm AIC}(k')>0,
\end{equation}
i.e., $f_{\rm AIC}(k)$ is a monotonic decreasing function. Hence, we
have
\begin{equation}\label{eqn:Append2-5}\nonumber
   \lim_{N\rightarrow \infty}\frac{{\rm AIC}(0)-{\rm AIC}(1)}{N}=
   2\log \frac{\left(\frac{1}{p}\sum^p_{i=1}l_i \right)^{p}}
     {\left(\frac{1}{p-1}\sum^p_{i=2}l_i\right)^{p-1}l_1}
     + \lim_{N\rightarrow \infty}\frac{-4p+2}{N}>0.
\end{equation}
If $N$ is finite but larger enough, we claim that Lemma 1 holds with
high probability. The high probability is also contributed by the
fact that, due to the property of SVD decomposition technique, the
first eigenvalue $l_1$ is always much larger than other eigenvalues.
Therefore, $2N\log \frac{\left(\frac{1}{p}\sum_{i=1}^p{l_i}
\right)^{p}} {\left(\frac{1}{p-1}\sum_{i=2}^p{l_i}\right)^{p-1}l_1}$
is larger enough to make Lemma 1 succeed at hypothesis ${\cal H}_1$.
Thus, we complete the proof of Lemma 1.

\section{Proof of Proposition 2}
\label{prof_pro3}
We firstly derive the derivation of the probability of misdetection $P_m$
(the probability for misdetecting the presence of primary user at hypothesis $H_1$),
then obtain the probability of detection $P_d$ through $1-P_m$.
Without loss of generality, the following derivation is also based on AIC.
According to \eqref{eqn:PdAIC}, we have
\begin{equation}\label{eqn:Append-c-Pm}
     {P_{m-AIC|H}} = {\rm Pr}[AIC(0)-AIC(1)<0|{\cal H}_1].\nonumber
\end{equation}
Similar to the process described in the proof of Proposition 1, we can rewritten $P_{m-AIC|H}$ as
\begin{equation} \label{eqn:Append-c-pmfull}
    {P_{m-AIC|H}} = {\rm Pr} \left( {\log}\left[\frac{({\frac{1}{p}}{\sum^p_{i=1}}{l_i})^p}
    {({{\frac{1}{p-1}}{\sum^p_{i=2}{l_i}}})^{p-1}l_1} \right]  < \frac{4p-2}{2N} \bigg|{\cal H}_1\right ).
\end{equation}
Where $\{l_1,l_2,\ldots,l_p\}$ are the decreasing ordered eigenvalues of the sampled covariance matrix ${\bf R}_x$ in \eqref{eqn:Rxn}.
When the number of observation $N$ is larger enough, we obtain the approximation
\begin{equation}\label{eqn:Append-c-expassume}
    \frac{1}{N}\sum^N_{i=1}{{\bm x}_i{\bm x}^{\dagger}_i} \approx {\cal E}\left({{\bm x}_i{\bm x}^{\dagger}_i}\right)
    = {\bf H}{\bf R}_s{\bf H^{\dagger}}+\sigma^2{{\bf I}_p}.\nonumber
\end{equation}
Thus
\begin{equation}\label{eqn:Append-c-lisumassme}
    \frac{1}{p}\sum^p_{i=1}{l_i} \approx {\frac{1}{p}}{\rm Tr}\left({\bf H}{\bf R}_s{\bf H^{\dagger}}\right)+\sigma^2.\nonumber
\end{equation}
Hence, \eqref{eqn:Append-c-pmfull} turns to
\begin{eqnarray}\label{eqn:Append-c-pm3}
    {P_{m-AIC|H}} \approx {\rm Pr}\left[{\frac{l_1}{\epsilon}}{\left(p-\frac{l_1}{\epsilon}\right)^{p-1}}
    >\frac{(p-1)^{p-1}}{\exp\big(\frac{2p-1}{N}\big)} \bigg|{\cal H}_1\right]
\nonumber\\
    = {\rm Pr}\left[y^p-py^{p-1}+\frac{(p-1)^{p-1}}{\exp\left(\frac{2p-1}{N}\right)}<0 \bigg|{\cal H}_1\right],
\end{eqnarray}
where $\epsilon={\frac{1}{p}}{\rm Tr}\big({\bf H}{\bf R}_s{\bf H}\big)+\sigma^2$ and $y \triangleq p-\frac{l_1}{\epsilon}$.

Assuming $\pi_1$ and $\pi_2$ (with $\pi_1<\pi_2$) are two real roots within $(0,p)$ of the following function
\begin{equation}\label{eqn:Append-c-jfun}
    g(y)=y^p-py^{p-1}+\frac{(p-1)^{p-1}}{\exp\big(\frac{2p-1}{N}\big)}.
\end{equation}
As described in the proof of proposition 1, the probability of misdetection is concluded as
\begin{equation}\label{eqn:Append-c-pm4}
    {P_{m-AIC|H}} \approx {\rm Pr}{\left[\pi_1<y<\pi_2 |{\cal H}_1\right]},\nonumber
\end{equation}
i.e.,
\begin{equation}\label{eqn:Append-c-pm5}
    {P_{m-AIC|H}} \approx {\rm Pr}{\left[(p-\pi_2)\epsilon<l_1<(p-\pi_1)\epsilon |{\cal H}_1\right]}.
\end{equation}
Note that $l_1$ is the largest eigenvalue of the sampled variance matrix ${\bf R}_x$.
Given the channel matrix, ${\bf R}_x$ can be approximated as
\begin{equation}\label{eqn:Append-c-approxiamtion}
    {\bf R}_x
    \approx {\frac{1}{N}}\left[{\bf H}{\sum^N_{i=1}{{\bm s}_i{{\bm s}_i}^{\dagger}}}{\bf H}^{\dagger}\right]+{\frac{1}{N}}{\sum^N_{i=1}{{\bm \mu}_i{{\bm \mu}_i}^{\dagger}}}
    \approx {\bf H}{{\bf R}_s}{\bf H}^{\dagger}+{\frac{1}{N}}{\sum^N_{i=1}{{\bm \mu}_i{{\bm \mu}_i}^{\dagger}}},\nonumber
\end{equation}
when $N$ is larger enough.

Let $\{\delta_1, \delta_2,\ldots, \delta_p \}$ and $\{\chi_1,\chi_2,\ldots,\chi_p \}$ be the
decreasing ordered eigenvalues of
${\bf H}{{\bf R}_s}{\bf H}^{\dagger}$
and ${\frac{1}{N}}{\sum^N_{i=1}{{\bm \mu}_i{{\bm \mu}_i}^{\dagger}}}$ respectively.
Apply Weyl's inequality theorem in \cite{Bhatia97},
the largest eigenvalue of ${\bf R}_x$, $l_1$, satisfies
\begin{equation}\label{eqn:Append-c-weyl}
    \chi_1+\delta_p \leqslant l_1 \leqslant \chi_1+\delta_1,\nonumber
\end{equation}
Equivalently $\chi_1$ satisfies
\begin{equation}\label{eqn:Append-c-chi1}
    l_1-\delta_1 \leqslant \chi_1 \leqslant l_1-\delta_p.
\end{equation}
Therefore,
there must exist a constant $\rho$ satisfying $\delta_p \leqslant \rho \leqslant \delta_1$ which makes $l_1-\rho$ equal to $\chi_1$.
Then \eqref{eqn:Append-c-pm5} is rewritten as
\begin{equation}\label{eqn:Append-c-pm6}
    {P_{m-AIC|H}} \approx {\rm Pr}{\left[(p-\pi_2)\epsilon-\rho<\chi_1<(p-\pi_1)\epsilon-\rho |{\cal H}_1\right]},\nonumber
\end{equation}
i.e.,
\begin{equation}\label{eqn:Append-c-pd1}
    {P_{d-AIC|H}} \approx {\rm Pr}{\left[\frac{(p-\pi_1)\epsilon-\rho}{\sigma^2}<\frac{\chi_1}{\sigma^2}<p |H_1\right]}+{\rm Pr}{\left[0<\frac{\chi_1}{\sigma^2}<\frac{(p-\pi_2)\epsilon-\rho}{\sigma^2} |{\cal H}_1\right]},\nonumber
\end{equation}
where we use the similar constraint for $\frac{\chi_1}{\sigma^2}$ as
in the proof of Proposition 1. Since $\chi_1$ converges to the
Tracy-Widom distribution of order two, we conclude
\begin{equation}\label{eqn:Append-c-Pd}
    P_{d-AIC|H}\approx Q(\rho),\nonumber
\end{equation}
where $Q(\cdot)$ is defined in Proposition 2.
Simultaneously, based on \eqref{eqn:Append-c-pm5}, the upper and lower bounds for $P_{m-AIC|H}$ is
\begin{equation}\label{eqn:Append-c-bounds}
    1-Q(\delta_1) \leqslant P_{m-AIC|H} \leqslant 1-Q(\delta_p).\nonumber
\end{equation}
Therefore, the upper and lower bound of $P_{d-AIC|H}$ can be obtain straightforwardly as
\begin{equation}\label{eqn:Append-c-boundsofPd}
    Q(\delta_p) \leqslant P_{d-AIC|H} \leqslant Q(\delta_1).\nonumber
\end{equation}
The proof for MDL criterion is the same, except that the function $g(y)$ in \eqref{eqn:Append-c-jfun}
is redefined as
\begin{equation}\label{eqn:Append-c-jfun2}
    g(y)=y^p-py^{p-1}+\frac{(p-1)^{p-1}}{\exp\big(\frac{(p-0.5)\log{N}}{N}\big)}.
\end{equation}
Proposition 2 is thus proved.

\bibliographystyle{IEEEtran}
\bibliography{IEEEabrv,Revision_version}

\begin{threeparttable}

\centering
\caption{Probability of false alarm with different $p=M\times K$ at $N=10000$}\label{TABLE:ChangeMKL}
\begin{tabular}{lrrrrr}
\toprule
                             &   $6=2\times3$    &   $12=3\times4$    &   $20=5\times4$    &   $24=4\times6$ &   $35=5\times7$\\
\midrule
Simulation results for SITC-AIC   &   0.0948  &   0.0770  &   0.0594  &   0.0541 &   0.0460\\
Simulation results for OITC-AIC   &   0.0972  &   0.0773  &   0.0597  &   0.0541 &   0.0470\\
Analytical results for SITC-AIC         &   0.1360  &   0.1036  &   0.0791  &   0.0711 &   0.0550\\
\midrule
Simulation results for SITC-MDL   &   0  &    0   &   0  &   0  &   0\\
Simulation results for OITC-MDL   &   0  &    0   &   0  &   0  &   0\\
Analytical results for SITC-MDL         &   0  &     0  &   0  &   0  &   0 \\
\bottomrule \end{tabular}
\end{threeparttable}
\
\

\begin{threeparttable}

\centering

\caption{Probability of false alarm with different $N$ at $p=MK=20$}\label{TABLE:ChangeN}
\begin{tabular}{lrrrr}
\toprule
                             &   $N=1000$    &   $N=5000$    &   $N=10000$  \\
\midrule
Simulation results for SITC-AIC   &   0.0421  &   0.0558  &   0.0594  \\
Simulation results for OITC-AIC   &   0.0421  &   0.0561  &   0.0597  \\
Analytical results for SITC-AIC   &   0.0581 &   0.0744 &   0.0791  \\
\midrule
Simulation results for SITC-MDL   &   0  &   0  &   0  \\
Simulation results for OITC-MDL   &   0  &   0  &   0  \\
Analytical results for SITC-MDL   &   0  &   0  &   0  \\
\bottomrule \end{tabular}
\end{threeparttable}

\end{document}